\documentclass[11pt]{article} 

\usepackage{amsmath,amsthm,latexsym,amssymb,amsfonts,epsfig}

\newtheorem{Theorem}{Theorem}[section]
\newtheorem{Definition}{Definition}[section]

\newcommand{\be}{\begin{equation}}
\newcommand{\ee}{\end{equation}}
\newcommand{\ba}{\begin{eqnarray}}
\newcommand{\ea}{\end{eqnarray}}

\title{
{\sf 
Renormalisation, wavelets and the Dirichlet-Shannon kernels}}
\author{
{\sf T. Thiemann}$^1$\thanks{{\sf 
thomas.thiemann@gravity.fau.de}}\\
\\
{\sf $^1$ Inst. for Quantum Gravity, FAU Erlangen -- N\"urnberg,}\\
{\sf Staudtstr. 7, 91058 Erlangen, Germany}\\
}
\date{{\small\sf \today}}

\makeatletter
\@addtoreset{equation}{section}
\makeatother

\begin{document} 

\maketitle

{\sf

\begin{abstract}
In constructive quantum field theory (CQFT) it is customary to first 
regularise the theory at finite UV and IR cut-off. Then one first removes 
the UV cutoff using renormalisation techniques applied to families of 
CQFT's labelled by finite UV resolutions and then takes the 
thermodynamic limit. Alternatively, one may try to work 
directly without IR cut-off.

More recently, wavelets have been proposed to define the renormalisation flow
of CQFT's which is natural as they come accompanied with a multi-resolution 
analysis (MRA). However, wavelets so far have been mostly studied in the 
non-compact case. Practically useful wavelets that display compact support
and some degree of smoothness can be constructed on the real line using 
Fourier space techniques but explicit formulae as functions of position are 
rarely available. Compactly supported wavelets can be periodised by 
summing over period translates keeping orthogonality properties but still
yield to rather complicated 
expressions which generically lose their 
smoothness and position locality properties.

It transpires that a direct approach to wavelets in the compact case is 
desirable. In this contribution we show that the Dirichlet-Shannon kernels 
serve
as a natural scaling function to define generalised
orthonormal wavelet bases on tori or copies of real lines respectively. 
These generalised wavelets are smooth, are simple explicitly computable 
functions, display quasi-local properties close 
to the Haar wavelet and have compact momentum supprt. Accordingly 
they have a built-in cut-off both in position and momentum, making them
very useful for renormalisation applications.
\end{abstract}

\section{Introduction}
\label{s1}

Examples of rigorously defined interacting Wightman quantum field theories 
in four dimensional Minkowski space are still not available.
As the usual perturbative approach is mathematically ill-defined 
(Haag's theorem) \cite{1}, the non-perturbative constructive QFT (CQFT)
programme 
was proposed \cite{2} which had spectacular success in two and three
spacetime dimensions \cite{3}. 
CQFT has both a rigorous path integral (Euclidian) 
and Hamiltonian formulation which are connected by Osterwalder-schrader 
reconstruction. The systematic construction of interacting QFT via the 
CQFT approach is to consider a family of theories labelled by a UV and IR 
cutoff. This could be a lattice spacing $M^{-1}$ and a toroidal radius 
radius $R$ respectively. Following the Wilsonian approach to renormalisation  
one first constructs a renormalisation flow defined by integrating out 
degrees of freedom at higher scales $M'>M$ to define an effective theory 
at scales $M$. Fixed points of this flow define consistent continuum theories  
at finite IR cutoff $R$ in the sense that the continuum theory 
which corresponds to infinite resolution $M\to \infty$ analysed at 
resolution $M$ coincides with the effective theory for the fixed point family.
In a non-trivial second step one then tries to take the thermodynamic 
limit $R\to \infty$.

In this work we are mainly but not only 
concerned with the renormalisation process at finite
IR cut-off, thus we consider QFT's at fixed finite $R$. It is convenient 
to study the QFT compactified on a d-torus $T^d$ and after trivial rescalings
of the coordinates we can restrict to the unit torus $T^d=[0,1)^d$. As 
renormalisation for all $d$ directions is done independently we can restrict
the considerations that follow to $d=1$. 

To define renormalisation, one has to specify what one means by 
``the theory at resolution M''. As quantum fields are operator valued 
distributions it is necessary to smear them with test functions, thus one can 
introduce the finite resolution scale $M$ by a suitable space of test 
functions $V_M$ labelled by $M$. The label set $\cal M$ from which the labels 
$M$ are taken is supposed to be equipped with some partial order $\le$ so that 
$M\le M'$ distinguishes between lower ($M$) and higher ($M'$) resolution. Then 
for $M\le M'$ the spaces $V_M$ and $V_{M'}$ are supposed to be nested i.e. 
$V_M\subset V_{M'}$ which means that the quantum field degrees of freedom 
at scale $M$ can be written as functions of the quantum field degrees of 
freedom at scale $M'$. This enables to integrate out the extra degrees
of freedom smeared by the functions in $V_{M'}-V_M$ and thus defines a 
renormalisation flow.

It is clear that the details of the flow depend on the choice of the spaces
$\{V_M\}_{M\in {\cal M}}$. 
However, the possible fixed point theories should not because 
its degrees of freedom can be smeared with any test functions and thus 
give ``cylindrically'' consistent effective theories labelled by the 
respective choice of test functions in $V_M$. As the above nested structure 
suggests, a systematic and ``economic'' approach to a suitable choice of the 
spaces $V_M$ are multi-resolution analyses (MRA's). Here the spaces 
$V_M$ are constructed from a single ``scaling'' test function $\phi$ whose 
rescaling by $M$ and translates provide an orthonnormal basis of $V_M$ with 
respect
to an inner product on the space $V$ of all test functions which thus carries 
a Hilbert space structure. The spaces $V_M$ are, however, by construction 
not mutually orthogonal but rather nested. To provide an orthonormal basis 
of $V$ one can thus construct a sequence of spaces inductively defined 
by $W_{M_0}:=V_{M_0}$ and $W_{M_{n}}$ is the orthogonal complement of 
$V_{M_n}$ in $V_{M_{n+1}}$. Here $n\mapsto M_n$ is a divergent, linearly 
ordered sequence in ${\cal M}$, i.e. $M_n<M_{n+1}$. This provides an
orthogonal decomposition $V=\oplus_{n=-\infty}^\infty W_{M_n}$
in the non-compact case and $V=\oplus_{n=0}^\infty W_{M_n}$
in the compact case. In discrete 
wavelet theory one often uses the seqence of scales $M_n=2^n\; M_0,\;
M_0=1$. 
A ``mother wavelet'' $\psi$ is now a very special test function namely its 
rescaling by $M_n$ and translates generate an orthonormal basis of $W_n$
and thus in turn of all of $V$. Given certain conditions on the scaling 
function $\phi$, the mother wavelet $\psi$ can be constructed from $\phi$
by Fourier analysis.  

Wavelet theory is an active research field of mathematics, mathematical 
physics and signal processing \cite{5}. In contrast to the ``plane wave'' 
basis used in the Fourier transform, wavelets by construction also display 
some notion of position space locality. One distinguishes between discrete 
wavelets (with discrete labels) and continuous wavelets (with continuous 
labels). The ones that naturally fit into the renormalisation language 
developed above are the discrete ones. Historically the first discrete wavelet 
was the Haar wavelet \cite{6} on the real axis whose scaling function 
is a step function. It is the only wavelet on 
the real axis known to date, whose dependence on position $x$ is known 
in closed form and which is of compact support. If one is content with
only quasi-compact support then the Shannon wavelet \cite{7}
which decays only 
slowly at infinity 
is an option if one is interested in explicitly known position space 
dependence (the scaling function is basically the sinc function). 
If manifest compact support is more important and in addition 
some degree of smoothness is required (the Haar wavelet is not even 
continuous) then one is lead to the Daubechies \cite{8} and Meyer
\cite{9} wavelets. It is well known 
that there is no ``Schwartz'' wavelet on the real axis, i.e. a wavelet 
which belongs to 
the space of Schwartz functions (smooth of rapid decrease) \cite{9}.

These well studied examples on the real axis generalise immediately to 
$\mathbb{R}^d$ using the tensor product. To obtain wavelets on compact 
spaces such as tori or spheres one can consider compactly supported children 
wavelets $\psi_{n,m}(x)=2^{-n/2}\; \psi(2^n x-m),\;m,n\in \mathbb{Z}$ 
on the real axis and periodise them by 
$[\pi\cdot\psi]_{n,m}(x)=\sum_{l\in\mathbb{Z}} \psi_{n,m}(x-l)$
which confines $n$ to $\mathbb{N}_0$ and $m=0,1,2,..,2^n-1$. This yields
an ONB of $L=L_2([0,1),dx)$ of periodic functions thanks to the compact 
support of 
$\psi$ but there are several 
drawbacks:\\
1. even if $\psi$ has compact support on $\mathbb{R}$ the support 
of $\pi\cdot \psi$ on $[0,1)$ 
may not even be quasi-loacal (i.e. the function is not peaked).\\
2. The complictated coordinate expression of $\psi$ propagates to 
$\pi\cdot \psi$.\\
3. Generically, the periodised wavelet is again discontinuous.\\
\\
It transpires that a more direct approach to wavelets on compact spaces such 
as intevals or circles is desirable which does not rely at all on the 
theory of the infinite line (see \cite{10a} for such real line 
based approaches which retain smoothness properties but are technically 
very involved). There are constructions available 
in the literature \cite{10} which use wavelet bases of finite order 
i.e. there is a maximal resolution $M_{\rm max}$ allowed. 
However, these are neither
peaked in position 
nor do they span all of $L_2$, they just span $V_{M_{\rm max}}$. 

We will understand MRA, scaling functions and wavelets in a generalised sense
which is inspired by minimal requirements that these should satisfy for 
purposes of renormalisation. These are \\
I. A nested sequence of subspaces $V_M\subset V_{M'},\; M\le M'$ whose span 
is dense in the Hilbert space $L$ of test functions. This allows to 
consider arbitrarily high resolutions and coarse grainings between different
resolutions.\\
II. A real valued orthonormal basis of test functions of $V_M$ which are 
obtained from a fixed finite set of ``mother'' (scaling) functions $\phi$ 
by rescaling and translation. Real valuedness is important because these 
functions are used for discretisations of quantum fields in the CQFT approach 
and we do not want to change their adjointness relations.\\
III. An orthonormal basis of $W_{M'}=V_M^\perp$ which is the orhogonal 
complement of $V_M$ in $V_{M'},\; M\le M'$ which is obtained 
from a fixed finite set of ``mother'' wavelet functions $\psi$, 
which are directly 
related to the scaling functions $\phi$, 
by rescaling and translation. The fact that the coarse graining maps 
built from the ONB of the $V_M$ or $W_M$ 
are based on a few scaling functions or wavelets makes the renormalisation
procedure systematic, economic and tractable.\\     
IV. The ONB should display at least peakedness in position space in order
that it can be used for discretisation of quantum fields on the lattice 
defined by the UV regulator.\\ 
\\
In this paper we show that such a generalised MRA on $T^1=S^1=[0,1)$ 
can be constructed 
based on the Dirichlet kernel \cite{13a}
as scaling function. It has the following 
features:\\
1. real valuedness.\\
2. smoothness.\\
3. compact momentum support.\\
4. it is a simple trigonometric polynomial which can be explicitly summed 
to obtain a simple position space expression which is related to the 
Shannon scaling function.\\
5. peakedness (quasi-locality) in position space.\\
6. its rescalings and translates generate an MRA.\\
7. There are two associated mother wavelets whose rescalings and translates
generate and ONB of $L_2([0,1),dx)$.\\
8. Being smooth, and reflection symmetric, it has an infinite number 
of vanishing trigonometric moments (the moments must 
be defined using trigonometric rather than proper polynomilas as the latter
are not periodic).\\
9. The wavelet basis can be considered as a smoothened version of the 
Haar wavelet on the torus with improved features for purposes of
renormalisation: not only field operators can be systematically discretised 
but also their derivatives (these are ill-defined in the discontinuous Haar
case).\\
\\
The locality features of this MRA is of course not surprising 
because it is well known that the rescalings of the Dirichlet kernel
provide smooth approximants of the periodic $\delta-$distribution.
However, to the best of our knowledge, the usefulness of the Dirichlet 
kernel for purposes of renormalisation and its relations to MRA's on 
$S^1$ have not been highlighted before. In that respect, the purpose of the 
present paper is to assmble available knowledge about the analytic 
properties of the Dirichlet kernel together with MRA and renormalisation 
framework. In tandem, we show that the Shannon kernel
on the real line, which we study from 
the above generalised point of view, has very similar properties.\\
\\
The architecture of this article is as follows:\\
\\

In section \ref{s2}, for the benefit of the unfamiliar reader we give a 
minimal account on MRA's and wavelets. This has the whole purpose of 
preparing for the next section and will be far from complete.

In section \ref{s3} we briefly recall what we mean by 
Hamiltonian renormalisation in the language of \cite{11} which has been 
applied and tested in \cite{11a} for free field theories without constraints 
and in \cite{11b} with constraints whose algebra is isomorphic to that 
of quantum gravity.  
We exhibit how coarse graining or blocking maps that define 
renormalisation flows are naturally 
generated by MRA structures. In particular we show that the renormalisation
flow in the works \cite{11a} is simply based on the Haar scaling function, 
which the authors of \cite{11a} were not aware of. Rather, the blocking 
kernels used there were obtained by rather independent arguments, 
specifically lattice gauge theory technology \cite{12}. For earlier
uses of MRA structures in the CQFT programme see e.g. \cite{13} and 
references therein.  
In \cite{11b} it became obvious that the renormalisation flow should 
be driven by 
kernels that display at least a minimal amount of smoothness which 
therefore directly motivated the present work. The impact of the choice 
of kernel on the physical properties of the fixed point theory was emphasised
before in \cite{11c}. 

In section \ref{s4} we define the Dirichlet and Shannon 
kernel, recall some of its 
analytical properties and demonstrate how it generates a generalised MRA and
an associated orthornormal mother wavelet pair. We highlight in what sense 
the corresponding blocking kernels can be considered as smooth versions of 
the Haar blocking kernel which makes it well adapted to discretisation 
of continuum QFT in the CQFT approach.  
 
In section \ref{s5} we showcase how the Dirichlet renormalisation flow 
tremendously simplifies the Haar flow of \cite{11a} while not changing
the fixed point theory. This is due to the translation invariance 
of both the Shannon and Dirichlet kernel which is 
not shared by the Haar kernel.    

In section \ref{s6} we summarise and conclude.

\section{Generalised Multi-Resolution Analysis (MRA)}
\label{s2}

We consider first the torus $X=T^1$ to define a generalised 
 MRA and after that explain where the definition has to be modified for 
the real line $X=\mathbb{R}$.\\ 
\\
We consider the torus $T^1$ as $\mathbb{R}/\mathbb{Z}$ i.e. as the interval 
$[0,1)$ with boundary points identified. By $L:=L_2([0,1),dx)$ we denote the 
the square integrable periodic functions on $[0,1)$. It has the orthonormal 
basis (ONB)
\be \label{2.1}
e_n(x):=e^{2\pi\;i\;n\; x},\;\;n\in \mathbb{Z}
\ee

We consider a subset ${\cal M}\subset \mathbb{N},\;1\in \{\cal M\}$ 
equipped with a partial 
order $\le$ i.e. an anti-symmetric, reflexive and transitive 
relation on $\cal M$ with respect to which it is also directed i.e. for any 
$M_1,M_2\in {\cal M}$ we find $M_3\in {\cal M}$ such that $M_1,M_2\le M_3$. 
We require that for pairs $M\le M'\in {\cal M}$ there is a scale factor number 
$s(M,M')\in \mathbb{N}$ and we define for $f\in L, s\in \mathbb{N}$ 
the dilatated function $(D_s f)(x)=f(s x)$ which is again 1-periodic. 
Furthermore, for any $M\in {\cal M}$ we require that there exists a 
dimension number $d(M)\in \mathbb{N}$ and for any $f\in L, d>0$ we define the 
translation $(T_d f)(x)=f(x-d)$. 
\begin{Definition}[Definition 2.1] \label{def2.1} ~\\
I.\\
A generalised 
multi-resolution analysis (MRA) of $L=([0,1),dx)$
subordinate to ${\cal M},\; s, d$ 
is an 
assignment ${\cal M}\ni M\mapsto V_M$ (principal translation invariant 
subspaces) where $V_M$ is a closed, finite
dimensional subspace of 
$L$ of dimension $d(M)$ such that\\
i. $V_1=\mathbb{C}$\\
ii. if $M\le M'$ then $V_{M}\subset V_{M'}$\\
iii. $\cup_{M\in {\cal M}} \; V_M$ is dense in $L$\\
iv. if $f\in V_M,\;M\le M'$ then $D_{s(M,M')}\; f\in V_{M'}$.\\
v. There exists a fixed finite set of scaling functions $\phi\in L$ such
that an ONB $\chi^M_m$ of $V_M$ is obtained as a fixed set of rational 
functions of their dilatations 
$D_{d(M)} \phi$
and translations $T_{1/d(M)}^m \phi,\;m=0,1,..,d(M)-1$ or combinations 
thereof.\\     
II.\\
A wavelet subordinate to a generalised MRA is a fixed finite set of functions 
$\psi\in L$ which are algebraic functions of the scaling functions $\phi$
such that a fixed set of
rational functions of their dilatations 
$D_{d(M)} \psi$
and translations $T_{1/d(M)}^m \psi,\;m=0,1,..,d(M)-1$ or combinations 
thereof provides an ONB of $W_M$ where $W_M=V_M^\perp$ is the orthogonal
complement of $V_M$ in $V_{M'(M)}$ and where $M'(M)\ge M$ is a fixed resolution
higher than $M$.     
\end{Definition}
A couple of remarks are in order:\\
1. In the usual wavelet literature on the real line 
one considers mostly the set 
${\cal M}=\{2^N,\;N\in \mathbb{Z}\}$ of integer powers of two with the usual 
linear order $\le$ on real numbers. The reason why in our case positive 
powers are sufficient is that negative powers would produce 
a lattice spacing larger than the lattice itself and thus 
maps us out of the  
space of 1-periodic functions.\\
2. The reason why we consider more general 
partial orders is because we allow more positive integers than positive 
powers of two and we wish that for $M\le M'$ the lattice defined by the 
points $m/d(M),\; m=0,1,..,d(M)-1$ is a sublattice of      
the lattice defined by the 
points $m'/d(M'),\; m'=0,1,..,d(M')-1$. \\
3. On the real line the spaces $V_M$ are all infinite dimensional 
i.e. $d(2^N)=\infty$ for all $N\in \mathbb{N}$ 
and instead of $V_1=\mathbb{C}$ we have $\cap_N V_{2^N}=\{0\}$.\\
4. On the real line the function one usually resticts $s$ to 
$2^N,2^{N+k}$ in which 
case it takes the value $s=2^k$. \\
5. On the real line the functions 
\be \label{2.3}
\chi^N_m(x):=2^{N/2}\; \phi(2^N x-m),\;\; m\in \mathbb{Z} 
\ee
are an orthonormal basis for $V_{2^N}$ if $\chi^0_m$ is an ONB of $V_{2^0}$.
We had to modify this for two reasons: First the space 
$V_{2^0}$ is only one dimensional on $S^1$ while infinite dimensional on 
$\mathbb{R}$ and thus cannot serve to build a basis for the higher dimensional
spaces $V_M$. Second, the integer shifts of an 1-periodic 
function are trivial. Therefore we disentangled the simultaneous rescaling 
and shifting performed on the function $\phi$ in (\ref{2.3}) and allowed pure 
dilatations and pure shifts or combinations thereof in order to assemble 
$\chi^M_m$ as a rational function (i.e. a fraction of polynomials)
of those. In the standard case (\ref{2.3}) 
we only need one such function $\phi$, the rational aggregate formed from
it is just the function itself multiplied by a constant. While the more 
general construction of the $\chi^M_m$ is more complicated than in the 
standard case, it keeps the spirit of building the basis $\chi^M_m$ of 
``children'' functions  from a few ``mother'' functions $\phi$. We restrict
to rational functions in order to keep the expressions involved managable 
and because in the examples we have rational functions appear to be 
sufficient.\\
6. On the real line the relation between scaling function $\phi$ and 
wavelet $\psi$ is less direct: it starts with a function $m_0$ in 
Fourier space subject to a support (Cohen's) condition and a normalisation 
condition on its modulus squared. Then one defines the Fourier transform of 
$\phi$ as an infinite product of dilatations of $m_0$ and the Fourier 
transform of $\psi$ is a product of $m_0$, the Fourier transform of $\phi$ 
and a phase factor depending on momentum. Only in rare cases can one 
solve the Fourier integral in closed form to obtain an explicit position 
space expression. Our definition is again motivated by the essential idea
that the wavelet basis should arise from a few mother wavelets which are 
computable by a concrete formula from the scaling functions. In contrast 
to the non-compact case we do not provide a procedure for how to obtain 
$\psi$ from $\phi$ but we allow for more complicated (algebraic rather than 
rational) relations between those, again because also in the standard case 
the relation is more complicated and because in the examples we have 
an algebraic relation appears to be sufficient. The algebraic rather than 
rational functions issue can however avoided if one increases the number
of mother wavelets. Thus, this difference is not essential.\\
\\
We now spell out how the definition needs to be or can be 
modified in the non-compact case $X=\mathbb{R}^1$. First of all,
the label set $\cal M$ can be generalised to a subset of the positive 
rationals $\mathbb{Q}_+$
equipped with some partial order with respect to to which it is directed.
For $M\le M'$ the number $s(M,M')$ is still required to be a positive 
natural. The function $d(M)$ is supposed to take values in $\mathbb{Q}_+$
and no longer has the interpretation of a dimension. 
\begin{Definition}[Definition 2.2] \label{def2.2} ~\\
A generalised 
multi-resolution analysis (MRA) and wavelet 
of $L=(\mathbb{R},dx)$ subordinate to 
${\cal M},\; s, d$ is identical to the generalised MRA and wavelet for the 
case $L=L_2([0,1),dx)$ with the following modifications:\\
I.1: $\cap_{M\in {\cal M}}=\{0\}$.\\
I.v and II.: the translations $T^m_{1/d(M)}$ 
are now labelled by $m\in \mathbb{Z}$ and are not confined to $0,..,d(M)$.
\end{Definition} 
It will be helpful to test the definition against a well known example
which can be used both in the non-compact and the compact case. This 
is the Haar wavelet. Its mother scaling function is given by 
\be \label{2.4}
\phi(x):=\chi_{[0,1)}(x) 
\ee
where $\chi_{[a,b)}$ denotes the characteristic function of the clopen 
interval $[a,b)$. 

Consider first the non-compact case. Then the functions 
$\chi^M_m(x)$ are for the set ${\cal M}=\{2^N,\; N\in \mathbb{Z}\}$ with 
$M:=2^N$ and $m\in \mathbb{Z}$
\be \label{2.5}
\chi^M_m(x):=M^{1/2}\; \chi_{[0,1)}(M\; x-m) 
\ee
which have support in $[x^M_m,x^M_{m+1}),\;x^M_m:=\frac{m}{M}$ and 
$d(M)=M$.
They are indeed just simple dilatations of translations of the mother scaling
functions. As these 
partition the real line into intervals of length $M^{-1}$ they are orthogonal
at fixed $M$ with respect to the standard inner product on 
$L=L_2(\mathbb{R},dx)$
\be \label{2.6}
<\chi^M_m,\; \chi^M_{m'}>=\delta_{m,m'}
\ee
Their span at fixed $M$ defines a dense subset of the subspace $V_M$ of $L$.
As 
\be \label{2.7}
\chi^M_m=2^{-1/2}\;[\chi^{2M}_{2m}+\chi^{2M}_{2m+1}]
\ee  
obviously 
$V_M\subset V_{2M}$. That $\cup_M\; V_M$ is dense in $L$ follows for instance
from the way the Lebsgue measure is constructed as a Borel mesure. Thus 
indeed we obtain an MRA on the real line using (\ref{2.4}). The orthogonal 
decomposition 
$V_{2M}=V_M\oplus W_M$ can be done by direct methods in this case: Obviously
we have to assemble an ONB of $W_M$ from the ONB of $V_{2M}$ because $W_M$
is a subspace thereof and every basis function of $W_M$ must be orthogonal to 
each of the basis functions of $V_M$. In view of (\ref{2.7}) 
this leads to the natural choice
\be \label{2.8}
\psi^M_m(x)=2^{-1/2}
[\chi^{2M}_{2m}(x)-\chi^{2M}_{2m+1}(x)]=(2M)^{1/2} \;
\psi(2M x-2m)  
\ee
where 
\be \label{2.9}
\psi(x)=2^{-1/2}\;[\chi_{[0,1)}(x)-
\chi_{[0,1)}(x-1)]
\ee
is the corresponding mother wavelet. It is a linear 
(and therefore rational) aggregate of mother scaling functions. 
Finally, as the intersection of the spaces $V_M,\; M=2^N,\;N\ge N_0$ coincides
with the space of square integrable functions that can be expanded
in the basis $\chi^M_m$ which are piecewise constant on intervals of length
$M^{-1}$ we see that for $N_0\to \infty$ only $\{0\}$ results as there is 
no non-vanishing, square integrable, constant function on $\mathbb{R}$. 

Now consider the compact case. First of all $\chi_{[0,1)}(x)\equiv 1$ for 
$x\in [0,1)=T^1$ so indeed $V_1=\mathbb{C}$. Otherwise we may use the 
same functions (\ref{2.7}) but with $x\in [0,1)$ and $m$ restricted to 
$0,1,..,M-1$ to define an $M$-dimensional subspace $V_M$ of $L=L_2([0,1),dx)$ 
consisting of mutually orthonormal, periodic functions on $[0,1)$ exploiting
the fact that $\chi^M_m(x)$ drops to zero outside of its support. From here
on the construction of $\psi$ follows exactly the same steps as in the 
non-compact case, just obeying the respective finite ranges $x\in [0,1)$ and 
$m\in \{0,1,..,M-1\}$ and $M=2^N\ge 1$
for $\psi^M_m$. In particular we see that the example 
of the Haar wavelet fits into the definition (\ref{2.7}) for an MRA 
on the torus with $d(M)=M=2^N$,
$s(M,2^k M)=2^k$ and $M'(M)=2$.

\section{Hamiltonian renormalisation}
\label{s3}

In this section we will combine the renormalisation technology 
from \cite{11} with the MRA framework developed in the previous section.\\
\\
Again it will be sufficient to consider one coordinate direction as 
$T^d,\mathbb{R}^d$ 
are Cartesian products. Thus for simplicity we consider a bosonic, scalar 
quantum field $\Phi$ (operator valued distribution) with conjugate 
momentum $\Pi$ on $[0,1)$
with canonical commutation and adjointness relations (in natural units 
$\hbar=1$) 
\be \label{3.1}
[\Pi(x),\Phi(y)]=i\; \delta(x,y),\;\;\Phi(x)^\ast=\Phi(x),\;\Pi^\ast(x)=\Pi(x)
\ee
where 
\be \label{3.2}
\delta(x,y)=\sum_{n\in \mathbb{Z}}\; e_n(x)\; e_n(y)^\ast,\; 
e_n(x)=e^{2\pi\;i\;n\;x}
\ee
is the periodic $\delta$ distribution on the torus or the standard $\delta$
distribution on the real line respectively. It is customary to 
work with the bounded Weyl operators for $X=[0,1)$ or $X=\mathbb{R}$ 
\be \label{3.2}
w[f,g]=\exp(i[\Phi(f)+\Pi(g)]),\;\;
\Phi(f)=\int_X\; dx\; f(x)\; \Phi(x),\;
\Pi(g)=\int_X\; dx\; g(x)\; \Pi(x)
\ee
with $f,g\in L=L_2(X,dx)$ test functions or smearing functions 
usually with some additional properties such 
as differentiability or even smoothness. For more complicated tensor fields 
or spinor fields a similar procedure can be followed (see e.g. the last 
two references in \cite{11a}).

Since the space $L$ enters the stage naturally we use MRA ideas to define 
a renormalisation group flow. Suppose that $M\mapsto V_M\subset$ with 
$M\in {\cal M}$ defines an MRA with orthogonal basis 
$\chi^M_m,\; m\in \mathbb{Z}_M=\{0,..,d(M)-1\}$ for $X=[0,1)$ and 
$\mathbb{Z}_M=\mathbb{Z}$ for $X=\mathbb{R}$ of $V_M$ normalised 
according to $||\chi^M_m||^2=d(M)^{-1}$. The reason why we do not normalise the 
$\chi^M_m$ (we could) will become obvious in a moment. In tandem with 
$V_M$ we define the space $L_M:=l_2(\mathbb{Z}_M)$ of square integrable 
sequences with $d(M)$ entries and with inner product
\be \label{3.3}
<f_M,g_M>:=d(M)^{-1}\;\sum_{m\in \mathbb{Z}_M} \; f^\ast_M(m)\; g_M(m)
\ee
This scalar product offers the interpretation of $f_M(m):=f(x^M_m),\; 
x^M_m:=\frac{m}{d(M)}$ and similar for $g_M$
as the discretised values of some 
functions $f,g\in L$ in which case (\ref{3.3}) is the Riemann sum 
approximant of $<f,g>_L$. It is for this reason that we did not normalise
the $\chi^M_m$. 

The spaces $V_M, L_M$ are in bijection via
\be \label{3.3}
I_M:\; L_M\to L,\;f_M\mapsto \sum_m\; f_M(m)\; \chi^M_m
\ee
Note that (\ref{3.3}) has range in $V_M\subset L$ only. 
Its adjoint $I_M^\dagger:\; L\to L_M$ is defined by 
\be \label{3.4}
<I_M^\dagger f,f_M>_{L_M}:=<f,\; I_M\; f_M>_L
\ee
so that 
\be \label{3.5}
(I_M^\dagger f)(m)=d(M)\; <\chi_M,f>_L
\ee
Clearly 
\be \label{3.6}
(I_M^\dagger I_M f_M)(m)=d(M)\; <\chi^M_m, I_M f_M>_L =f_M(m)
\ee 
i.e. $I_M^\dagger I_M=1_{L_M}$ while 
\be \label{3.7}
(I_M I_M^\dagger f)(x)=d(M)\; \sum_m\; \chi^M_m(x) <\chi^M_m,f_M>_L
=(p_M f)(x)
\ee
is the projection $p_M:\; L\mapsto V_M$.

Given $M\le M'$ we define the coarse graining map
\be \label{3.8}
I_{MM'}:=I_{M'}^\dagger \; I_M:\; L_M\mapsto L_{M'}
\ee
It obeys 
\be \label{3.9}
I_{M'}\; I_{MM'}=p_{M'}\; I_M=I_M
\ee  
because $I_M$ has range in $V_M\subset V_{M'}$ for $M\le M'$. 
This is the place where the MRA property of the nested set of subspaces 
$V_M$ was important. Next for $M_1\le M_2\le M_3$ we have 
\be \label{3.9}
I_{M_2 M_3}\; I_{M_1 M_2}=
I_{M_3}^\dagger\; p_{M_2}\; I_{M_1}=  
I_{M_3}^\dagger\; I_{M_1}=I_{M_1 M_3}
\ee
for the same reason. This is called the condition of cylindrical 
consistency which is crucial for the renormalisation group flow.

To see the importance of (\ref{3.9}) we consider a probability 
measure $\nu$ on the 
space $\cal F$ of field configurations $\Phi$ which defines a Hilbert space   
${\cal H}=L_2({\cal F},d\nu)$ and a representations space for the Weyl
algebra $\mathfrak{A}$ generated from  the Weyl elements (\ref{3.2}). We 
set $w[f]:=w[f,g=0]$ and define the generating functional of moments of $\nu$
by 
\be \label{3.10}
\nu(f):=\nu(w[f])
\ee
If we restrict $f$ to $V_M$ we obtain an effective measure on the space 
of discretised quantum fields $\Phi_M=I_M^\dagger \Phi$ via 
\be \label{3.11}
w[I_M f_M]=w_M[f_M]=e^{i\Phi_M(f_M)},\; \Phi_M(f_M)=<f_M,\Phi_M>_{L_M}
\ee
and 
\be \label{3.12}
\nu_M(f_M):=\nu(w[I_M f_M])=\nu_M(w_M[f_M])
\ee
The measures $\nu_M$ on the spaces ${\cal F}_M$ of fields $\Phi_M$ are 
consistently defined by construction
\be \label{3.13}
\nu_{M'}(I_{MM'} f_M)=\nu_M(f_M)
\ee
for any $M\le M'$ since the $\nu_M$ descend from a continuum measure.
Conversely, given a family of measures $\nu_M$ satisfying
(\ref{3.13}) a continuum measure $\nu$ can be constructed known as the 
projective limit of the $\nu_M$ under mild technical assumptions 
\cite{14}. To see the imprortance of (\ref{3.9}) for his to be the case,
suppose we write $f\in L$ in two eqivalent ways 
$f=I_{M_1} f_M=I_{M_2} g_{M_2}$ then we should have 
$\nu_{M_1}(f_{M_1})=\nu_{M_2}(g_{M_2})$. Now while $M_1,M_2$ may not be 
in relation, as $\cal M$ is directed we find $M_1,M_2\le M_3$. Applying 
$I_{M_3}^\dagger$ we conclude $I_{M_1 M_3} f_{M_1}=I_{M_2 M_3} g_{M_2}$ thus 
due to (\ref{3.13}) indeed
\be \label{3.14}
\nu_{M_1}(f_1)=\nu_{M_3}(I_{M_1 M_3} f_{M_1})=
\nu_{M_3}(I_{M_2 M_3} g_{M_2})=
\nu_{M_2}(g_{M_2})
\ee
In CQFT the task is to construct a representation of $\mathfrak{A}$ with 
additional properties such as allowing for the imlementation of a 
Hamiltonian operator $H=H[\Phi,\Pi]$ which imposes severe restrictions 
on the Hilbert space representation. One may start with discretised 
Hamiltonians ($H$ is the classical Hamiltonian function)
\be \label{3.15}
H^{(0)}_M:[\Phi_M,\Pi_M]:=H(p_M\Phi,p_M\Pi]
\ee
on ${\cal H}^{(0)}_M:=L_2({\cal F}_M,\nu^{(0)}_M)$ where $\nu^{(0)}_M$ is 
any probability measure to begin with, for instance a Gaussian measure
or a measure constructed from the ground state $\Omega^{(0)}_M$ of the 
Hamiltonian $H^{(0)}_M$. The definition (\ref{3.15}) is incomplete without
some ordering prescription, we assume that such a prescription has been 
chosen.

The point of using an IR cutoff, that is the compact space $X=[0,1)$,
is that there are only finitely 
many, namely $d(M)$ degrees of freedom $\Phi_M,\Pi_M$ which are conjugate 
\be \label{3.16}
[\Pi_M(m),\Phi(m')]=i\; d(M)\; \delta(m,m'),\;\;\Phi_M(m)^\ast=
\Phi_M(m),\;\Pi_M^\ast(m)=\Pi_M(m)
\ee
so that construction of $\nu^{(0)}_M$ does not pose any problems. For 
$X=\mathbb{R}$ (\ref{3.16}) still holds, but now existence 
of $\nu^{(0)}_M$ is not granted and requires further analysis. Assuming
this one 
fixes for each $M\in {\cal M}$ an element $M\le M'(M)\in {\cal M}$ and 
defines isometric injections 
\be \label{3.17}
J_{MM'(M)}:\; {\cal H}^{(n+1)}_M\to {\cal H}^{(n)}_{M'(M)},\;\;
{\cal H}^{(n)}_M:=L_2({\cal F}_M,d\nu^{(n)}_M)
\ee
via 
\be \label{3.18}
\nu^{(n+1)}_M(f_M):=\nu_{M'(M)}(I_{MM'(M)} f_M)
\ee
and with these the flow of Hamiltonians
\be \label{3.19}
H^{(n+1)}_M:=J_{MM'(M)}^\dagger\; H^{(n)}_{M'(M)}\; J_{MM'(M)}
\ee
The isometry of the injections relies on the assumption that the span of the 
$w_M[f_M]$ is dense in ${\cal H}^{(0)}_M$ which is typically the case. 

This defines a sequence or flow (indexed by $n$) of families (indexed by $M$) 
of theories ${\cal H}^{(n)}_M,H^{(n)}_M$. At a critical or fixed point of 
this flow the consistency condition (\ref{3.13}) is satisfied (at first 
in the linearly ordered sets of ${\cal M}(M):=\{(M')^N(M),\; N\in
\mathbb{N}_0\}$ and then ususally for all of $\cal M$ by universality) 
and one obtains a consistent family $({\cal H}_M,\;H_M)$. This family
defines a continuum theory $({\cal H}, H)$ as one obtains 
inductive limit 
isometric injections $J_M:\; {\cal H}_M \mapsto {\cal H}$ such that 
$J_{M'} J_{MM'}=J_M,\;M\le M'$ thanks to the fixed point identiy  
$J_{M_2 M_3}\;J_{M_1 M_2}=J_{M_1 M_3},\;M_1\le\; M_2\le M_3$ and such that
\be \label{3.20}
H_M=J_M^\dagger \; H\; J_M
\ee
is a consistent family of quadratic forms 
$H_M=J_{MM'}^\dagger\; H_{M'}\; J_{MM'},\; M\le M'$.\\
\\
The conclusion of the present section is that the MRA framework fits quite 
naturally with the construction of the Hamiltonian renormalisation flow.
All that is needed is in fact the nested structure of the $V_M$, it is 
strictly speaking not necessary to have mother scaling functions or 
mother wavelets. However, to reduce the arbitrariness in the nested structure
or choice of coarse graining maps,
the requirement that the nesting descends from a (finite number of)
scaling function(s) is very useful. In fact in \cite{11a} the authors used,
without being aware of it, the MRA based on the Haar scaling function. While 
this works, it makes the formalism unnecessarily complicated because 
the Haar scaling function is not even continuous and thus the discretisation
prescription (\ref{3.15}) is ill defined as it stands as soon as 
$H$ depends on derivatives of $\Phi,\Pi$ which is typically the case.
Thus in \cite{11a} one had to use an additional prescription to define 
those discrete derivatives which increases the discretisation ambiguity
which one actually wants to avoid. 
It is for this reason that we try to base an MRA on scaling functions 
with additional smoothness properties. We will show in the next section that 
one possibility is based on the Shannon and Dirichlet kernels.

\section{Resolutions of the identity MRA's and the Shannon-Dirichlet kernels}
\label{s4}

In the previous section we have shown that a nested structure of subspaces 
$V_M\subset L$ with $V_M\subset V_{M'},\; M\le M',\; M,M'\in {\cal M}\subset
\mathbb{N}$ whose span is dense in $L$ leads to a useful renormalisation
flow in CQFT for any choice of orthonormal basis 
$\chi^M_m,\; m=0,1,..,d(M)=\dim(V_M)$. This uses only part of the definition 
of an MRA: It was not specified how that basis of $V_M$ is to be generated,
in particular it was not required that the $\chi^M_m$ descend from one 
or several fixed mother scaling functions. To systemize this choice the 
concept of mother functions and therefore the full definition of an MRA 
appears natural.

\subsection{Torus}
\label{s4.1}

We begin with the compact case.
First of all, one may pick an ONB $e_n$ of L which for sake of defininiteness
we label by $n\in \mathbb{Z}$ (if one prefers $n\in \mathbb{N}_0$ 
set $b_{2n}:=[e_n+e_{-n}]/\sqrt{2};n\ge 0,\; 
b_{2n+1}:=[e_n-e_{-n}]/\sqrt{2},n>0$). This could be the eigenbasis of 
a self-adjoint operator on $L$ with pure point spectrum. For reasons
explained in section \ref{s3}, we want the $e_n$ to have at least some 
degree of differentiability. Then for any 
odd integer $M$ we may consider the $d(M)=M$ dimensional 
subspaces $V_M$ of $L$ spanned by 
the functions $e_n,\; |n|\le (M-1)/2$. Picking ${\cal M}\subset \mathbb{N}$ as 
the odd naurals equipped with the usual ordering relation $\le$ on the 
naturals, one obtains trivially a nested structure of Hilbert spaces. 

However, this is still too general and not useful for our renormalisation
inentions. This is because the $e_n(x)$ typically fail to be localised 
with respect to $x$, because the spectral label $n$ has in general nothing 
to do 
with the points $x^M_m=\frac{m}{M}$ of the lattice of $[0,1)$ at which 
we wish to localise and discretise our quantum fields $\Phi,\Pi$. Thus 
we need to connect the label $n$ to the lattice label $m$ in such a way 
that the resulting orthogonal basis functions $\chi^M_m$ display some 
form of peakedness in position space around the points $x^M_m$. 

To do this, we use the following notation: Let for $M$ odd
$\mathbb{Z}_M:=\{0,1,2,..,M-1\}$
and $\hat{\mathbb{Z}}_M:=\{-\frac{M-1}{2},-\frac{M-1}{2}+1,..,
\frac{M-1}{2}\}$ and $d(M):=M$. Pick a unitary $M\times M$ matrix 
with entries $e^M_n(m),\; n\in \hat{\mathbb{Z}}_M,\; m\in \mathbb{Z}_M$
and consider 
\be \label{4.1}
\chi^M_m(x):=M^{-1/2}\;
\sum_{n\in \hat{\mathbb{Z}}_M}\; e_n(x)\; [e^M_n(m)]^\ast     
\ee 
Then by construction
\be \label{4.2}
<\chi^M_m,\; \chi^M_{m'}>_L=M^{-1}\;\delta_{m,m'}
\ee
The question now arises whether it is possible to pick that unitary matrix 
in such a way that $\chi^M_m(x)$ 1. is real valued (so that they can 
used to define Weyl operators), 2. is localised around $x^M_m$ and 3.
such that 
the $\chi^M_m$ descend from some mother scaling functions $\phi$ in the 
generalised sense of section \ref{s2}.

We will not give an exhaustive answer about the maximal freedom there is 
in doing so but rather show that there is at least one example that satisfies 
all three criteria. Moreover, the resulting $\chi^M_m$ will not only be 
smooth (thus its Fourier coefficients decay rapidly at infinity) but even
trigonometric polynomials (i.e. finite linear combinations of the 
eigenbasis of the momentum operator $-i d/dx$ on $L$, hence the Fourier
coefficients are of compact support).\\ 
\\
We pick for $m\in \mathbb{Z}_M,\; n\in \hat{\mathbb{Z}}_M$
\be \label{4.3} 
e_n(x):=e^{2\pi\;i\;n\;x},\; e^M_n(m):=e_n(x^M_m),\; x^M_m:=\frac{m}{M}
\ee
Then indeed
\be \label{4.4}
\frac{1}{M}\; \sum_{m\in \mathbb{Z}_M}\; e^M_n(m)\;[e^M_{n'}(m)]^\ast
=\delta_{n,n'},\;
\sum_{n\in \hat{\mathbb{Z}}_M}\; e^M_n(m)\;[e^M_n(m')]^\ast=M\delta(m,m')
\ee
and we have, using $e_n^\ast=e_{-n},.\; e_n \; e_{n'}=e_{n+n'}$ 
\be \label{4.5}
\chi^M_m(x)=M^{-1}\;\sum_{n\in \hat{\mathbb{Z}}_M}\; e_n(x)\; 
[e_n^M(m)]^\ast=\sum_{|n|\le \frac{M-1}{2}}\; e_n(x-x^M_m)
\ee
from which the real valuedness of $\chi^M_m$ is manifest. Also clearly 
$\chi^M_m$ is smooth being a trigonometric polynomial of order $(M-1)/2$ 
and thus has compact momentum support rather than having only rapid momentum
decrease. Furthermore the geometric series (\ref{4.5}) can be explicitly 
summed to yield the explicit expression 
\be \label{4.6}
M\;\chi^M_m(x)=\frac{\sin(\pi\; M\;[x-x^M_m])}{\sin(\pi[x-x^M_m])}
\ee
which is a rational function of dilatations and translations of the 
$\sin$ function. Thus, if we define the {\it mother scaling} function to be 
\be \label{4.7}
\phi(x):=\sin(\pi\; x)
\ee
then 
\be \label{4.8}
M\;\chi^M_m(x)=\frac{[D_M\; T_{1/M}^m\;\phi](x)}{[T_{1/M}^m \phi](x)}
\ee
which is precisely of the form required in definition \ref{def2.1} if we 
remember that $d(M)=M$. To complete the definition we must decide 
on the choice of ${\cal M}$ and its partial order. We pick $\cal M$ to be 
the odd naturals and define $M\le M'$ iff $\frac{M'}{M}$ is a (necessarily
odd) integer. This partial order is motivated by the requirement that the 
lattice labelled by $M$ should be a sublattice of the lattice labelled by 
$M'$. With this partial order, $\cal M$ is directed as given $M_1,M_2$ 
we may pick $M_3=M_1 M_2$ 
(or more economically the smallest common multiple) to achieve 
$M_1, \; M_2\le M_3$.
  
Finally, we note that $M\chi^M_m(x)=\delta_M(x-y),\; y=x^M_m$ is the 
restriction to our lattice points $x^M_m$ of the {\it Dirichlet kernel}
\be \label{4.9}
\delta_M(x-y)=\sum_{|n|\le \frac{M-1}{2}}\; e_n(x)\; [e_n(y)]^\ast
\ee
which is an approximant to the $\delta-$distribution on $[0,1)$ cut off 
at momentum $(M-1)/2$. This makes it plausible to be strongly peaked at 
$x=x^M_m$ as $M$ grows large. To investigate this, we perform some elementary
analysis on the function $\delta_M$. Having period 1 and being symmetric 
around $x=0$ it will be sufficient to investigate the interval $x\in 
[0,\frac{1}{2})$ of the function 
\be \label{4.10}
\delta_M(x)=\frac{\sin(\pi\;M\;x)}{\sin(\pi\;x)}
\ee
whose denominator vanishes only at $x=0$ in $[-1/2,1/2)$. However $\delta_M$
is smooth at $x=0$ with $\delta_M(0)=M$.
Close to $x=0$ it becomes the sinc function $M\;{\rm sinc}(\pi\; M\; x)$ 
which is the scaling function of the Shannon wavelet on the real line.
Besides $x=0$ the numerator vanishes at the zeroes $z^m_m=\frac{m}{M},\; 
m=1,..,\frac{M-1}{2}$. To compute its extrema between those zeroes 
we take the derivative
\be \label{4.11}
[\delta_M]'(x)=\frac{\pi}{\sin^2(\pi x)}\;
[M\; \cos(\pi\; M\; x)\;\sin(\pi x)
-\sin(\pi\; M\; x)\;\cos(\pi x)]
\ee
It vanishes at $x=0$ as the numerator $\propto x^3$ while the denominator 
$\propto x^2$ there. It also vanishes at $x=1/2$ due to the cosines and 
because $M$ is odd. For $0<x<1/2$ both $\cos(\pi x),\sin(\pi x)$ are non 
vanishing and since a zero of $\cos(\pi M x)$ is an extremum of 
$\sin(\pi Mx)$   
the vanishing of (\ref{4.11}) for $0<x<1/2$ yields the transcendental equation
\be \label{4.12}
\tan(M\;y)=M\tan(y);\;\;0<y=\pi x<\frac{\pi}{2}
\ee
This equation has $1+\frac{M-1}{2}$ solutions $y^M_m,\; m=0,.,,
\frac{M-1}{2}$
with $0=y^M_0<z^M_1<y^M_2<..<z^M_{[M-1]/2}<y^m_{[M-1]/2}<\frac{1}{2}$.
To see this, note that the right hand side is positive, strictly monotonously 
increasing and diverges at $y=\frac{\pi}{2}+$. The left hand side runs through 
one positive fundamental branch of the $\tan$ function between $y=0,\pi/(2M)$,
and $(M-1)/2$ full (negative and positive) fundamental branches between 
$y=(2k-1)/(2M)\pi, (2k+1)/(2M)\pi,\; k=1,..,(M-1)/2$. Since $\tan(M\;y)$ is
strictly monotonously increasing but at a faster rate than $M\tan(y)$ in 
each of those full branches we get one solution. The solution 
$y^M_k,\; k=1,..,\frac{M-1}{2}$ lies very close to $\frac{2k+1}{2M}\pi$,
the larger $k$ (because $\tan(y)$ is monotonously increasing) and the larger 
$M$ (since $[\tan(M y)-M\tan(y)]'=M[\tan^2(My)-\tan^2(y)]>0$ for $y>0$
we have $\tan(My)>M\tan(y)$ for $0<y<\pi/(2M)$ and $d/dM
[\tan(My)-M\tan(y)=1+\tan^2(My)-\tan(y)>1+\tan^2(y)-\tan(y)
=[\tan(y)+1/2]^2+3/4>0$). Therefore we can construct them iteratively by 
setting $y^M_k=:\frac{2k+1}{2M}\pi-\Delta^M_k$ with $\Delta^M_k\le 
\frac{\pi}{M}$ and writing (\ref{4.12}) as 
\be \label{4.13}
\tan(M\Delta^M_k)=\frac{1}{M}\cot(y^M_k)
\ee
which grants that $\Delta^M_k<\pi/(2M)$. We can solve (\ref{4.13}) by 
reinserting it into itself. To lowest order in $1/M$
\be \label{4.14}
\Delta^M_k=\frac{1}{M}\arctan(\frac{\cot((2k+1)/(2M)\pi}{M})  
\ee
The value of $\delta_M$ at $y^M_k$ can be seen from 
\be \label{4.15}
[\delta_M(x=y^M_k/\pi)]^2=\frac{M^2}{1+[M^2-1]\sin^2(y^M_k)}
\ee
For $k=0$ we get $\delta_M^2(0)=M^2$ while for $k>0$ we get
$\delta_M^2(y^M_k/\pi)=O(1)$. Thus the maximum at $x=0$ exceeds the 
other extrema by at at least an order of $M$. 

Accordingly one can visualise $\delta_M$ 
roughly as smoothened version of a symmetric
triangle of height $M$ and width $2/M$ between the first zeroes 
$z^M_1=\pm 1/M$ of $\delta_M$. Outside that interval which has volume smaller
than $1$ the function is relatively bounded as $O(1/M)$ compared to its 
maximum. Thus the central triangle has area $O(1)$. If we compare to 
the Haar scaling function 
$M\chi^M_m(x)=M\chi_{[-\frac{1}{2M},\frac{1}{2M})}(x-m)$
we see that basically the rectangle of height $M$ and width $1/M$ has 
been replaced by that triangle, except for subdominant contributions 
to the triangle which are the price to pay for having a smooth kernel.

That price however is well worth paying for: We want to use $\delta_M$ to 
discretise functions $f$ on the lattice $x^M_m=\frac{m}{M}$ by 
\be \label{4.15a}
f_M(m):=[I_M^\dagger f](m)=M\;<\chi^M_m,f>_L
\ee
This formula is well defined 
for both the Haar and Dirichlet scaling function. However, what about 
derivatives? Using the same formula we would get 
$[f']_M(m)=<\chi^M_m,f'>$ which is still well defined but one would like 
to relate this to some sort of discrete derivative of $f_M$. In the case 
of the Haar scaling function one can do the integral and obtains 
$[f']_M(m)=M[f(x^M_{m+1})-f(x^M_m)]$ which is a possible definition of the 
discrete derivative, however, the function values $f(x^M_m)$ are {\it not
linearly}
related to the values $f_M(m)$ in (\ref{4.15a}). In particular, if we replace 
$f$ by the quantum field, this definition of derivative would map us out 
of the space of already discretised fields $\Phi_M(m)$.

By contrast, in the case of the Dirichlet
kernel, as the $\chi^M_m$ are smooth we can integrate by parts (no 
boundary terms occur because all functions involved are periodic)
to obtain 
\be \label{4.16}
[f']_M(m)=-<[\chi^M_m]',\;f>_L
\ee
The functions $\chi^M_m$ are in $V_M$ which is spanned by the 
$e_n,\;|n|\le (M-1)/2$. As $e_n'=2\pi\; i\; n\ e_n$ the functions 
$[\chi^M_m]'$ are still in $V_M$ and thus can be expressed as linear 
combinations of the $\chi^M_{m'}$. It follows that (\ref{4.16}) defines
a linear map on the sequence $m\mapsto f_M(m)=<\chi^M_m,f>_L$
\be \label{4.17}
[\partial_M f_M](m):=[f']_M(m)=\sum_{\tilde{m}\in \mathbb{Z}_M}\;
\partial_M(m,\tilde{m})\; f_M(\tilde{m})
\ee
Without 
working it out exlicitly, we can already determine the dominant contribution    
of the matrix $\partial_M$: Since $\chi^M_m$ is steepest of inclination 
$\pm M$ close to $x^M_{m \mp 1}$ we know without further calculation that 
$\partial_M(m,\tilde{m})$ will be approximated by 
$c\; M[\delta_{\tilde{m},m+1}-\delta_{\tilde{m},m-1}]$ where $c$ is a 
numerical constant of order unity. By construction, this is an antisymmetric 
matrix as being related to the derivative of a symmetric kernel.

Accordingly, the formula (\ref{4.15a}) can be {\it universally} 
used to discretise
fields, their momenta and their arbitrarily high derivatives as they appear
in the classical Hamiltonian without 
introducing extra structure, thereby downsizing the discretisation ambiguities.
All of the renormalisation programme can therefore be based on a {\it single}
input, namely the MRA based on a scaling function with suitable properties 
that we listed in the introduction.\\
\\
We complete the analysis by computing the mother wavelets corresponding to 
the Dirichlet scaling function. This requires, for each $M\in {\cal M}$,
the specification of a fixed
element $M'(M)\in {\cal M}$ such that $M'(M)>M$. We pick the simplest choice 
$M'(M)=3\; M$. We will content ourselves with considering the linearly 
ordered subset ${\cal M}_3;=\{3^N,\; N\in \mathbb{N}_0\}$ which is what 
one always does in discrete wavelet analysis (with $2$ replaced by $3$). 
Note that for renormalisation the essential structure is the MRA and 
mother scaling function while the mother wavelet function is a convenient
but not essential additional structure.

We have with $\delta^N:=\delta_{M=3^N}$ and since 
$(3^{N+1}-1)/2=3^N+(3^N-1)/2$
\be \label{4.18}
\delta^{N+1}=\sum_{|n|\le \frac{3^{N+1}-1}{2}}\; e_n
=(e_{3^N}+1+e_{-3^N})\;\sum_{|n|\le \frac{3^N-1}{2}}\; e_n
=(e_{3^N}+1+e_{-3^N})\;\delta^N
\ee
which displays a self-similar structure
\be \label{4.19}
\delta^N=\prod_{k=1}^{N-1}\;\gamma^k,\;\gamma^k=e_{3^k}+1+e_{-3^k}
\ee
and nicely illustrates how the Dirichlet kernel $\delta^N$ is built 
from the $3^N$ basis vectors $e_n,\; |n|\le (3^N-1)/2$. This also 
makes it easy to give an explicit parametrisation of the orthogonal 
complement $W_{3^N}$ of $V_{3^N}$ in $V_{3^{N+1}}$ which has twice 
the dimension of $V_{3^N}$ which is $d(3^N)=3^N$: it is given 
by the functions $e_{n\pm 3^N},\; |n|\le (3^N-1)/2$. Let us therefore 
define $e^\sigma_n:=e_{n+\sigma 3^N},\; |n|\le (3^N-1)/2,\; 
\sigma\in \{0,\pm 1\}$. Then 
\be \label{4.20}
<e^\sigma_n,\; e^{\tilde{\sigma}}_{\tilde{n}}>_L
=\delta_{\sigma,\tilde{\sigma}}\; \delta_{n,\tilde{n}}
\ee
We are thus lead to consider two kernels $\delta^N_\pm$ in addition to 
$\delta^N:=\delta^N_0$ which are also real valued
\be \label{4.21}
\delta^N_+(x)=\cos(3^N\;2\;\pi\;x)\;\delta^N(x),\;
\delta^N_-(x)=\sin(3^N\;2\;\pi\;x)\;\delta^N(x)
\ee
and set for $\sigma=\pm 1$
\be \label{4.22}
\psi^N_{\sigma,m}(x):=\delta^N_\sigma(x-x^N_m)
\ee
We have with $M=3^N$
\ba \label{4.23}
&& <\psi^N_{+,m},\;\psi^N_{-,\tilde{m}}>
= \frac{1}{4i}\; 
<(e_{3^N}\;e_{-3^N}(x^M_m)+e_{-3^N}\;e_{3^N}(x^M_m)\chi^M_m,
,\;(e_{3^N}\;e_{-3^N}(x^M_m)-e_{-3^N}\;e_{3^N}(x^M_m)\chi^M_{\tilde{m}}>
\nonumber\\
&=&
\frac{1}{4i}\;
[
e_{3^N}(x^M_m-x^M_{\tilde{m}})
-e_{-3^N}(x^M_m-x^M_{\tilde{m}})]\;,<\chi^M_m,\; \chi^M_{\tilde{m}}>
\nonumber\\
&=& 0
\ea
where we used that 
$e_{3^N}\chi^M_n$ contains only modes
$(3^N+1)/2\le 3^N+n\le (3^{N+1}-1)/2$ while
$e_{-3^N}\chi^M_n$ contains only modes
$-(3^N+1)/2\ge -3^N+n\ge -(3^{N+1}-1)/2$ 
which lie in mutually disjoint sets. On the other hand 
\ba \label{4.24}
&&<\psi^N_{\pm,m},\;\psi^N_{\pm,\tilde{m}}>
=\frac{1}{4}\; 
<(e_{3^N}\;e_{-3^N}(x^M_m)\pm e_{-3^N}\;e_{3^N}(x^M_m)\chi^M_m,
,\;(e_{3^N}\;e_{-3^N}(x^M_m)\pm e_{-3^N}\;e_{3^N}(x^M_m)\chi^M_{\tilde{m}}>
\nonumber\\
&=&
\frac{1}{4}\;
[
e_{3^N}(x^M_m-x^M_{\tilde{m}}))
+e_{-3^N}(x^M_m-x^M_{\tilde{m}}))]\;,<\chi^M_m,\; \chi^M_{\tilde{m}}>
\nonumber\\
&=& \frac{1}{2}\; 3^N\; \delta_{M,\tilde{m}}
\ea
Accordingly, the $\psi^N_{\pm,n},\;|n|\le (3^N-1)/2$ are an ONB for 
$W_{3^N}$ up to normalisation and as $N$ varies 
they provide and ONB of $L$. We now relate them 
to the scaling function using $M=3^N,\; y^M_m=x-x^M_m,\;m\in \mathbb{Z}_M$ 
and determine the mother wavelets. We have with normalisation $2M=2\;3^N$
(the dimension of $W_N$)
\ba \label{4.25}
\psi^N_{+,m}(x)
&=&
2\frac{\cos(2\pi \; M y^M_m)\;\sin(\pi\; M\;y^M_m)}{\sin(\pi\; y^M_m)}
=2\frac{\sin(\pi\; M\;y^M_m)-2\sin^3(\pi\; M\; y^M_m)}{\sin(\pi\; y^M_m)} 
\nonumber\\
&=& \frac{\sin(3\pi\; M\;y^M_m)-\sin(\pi\; M\; y^M_m)}{\sin(\pi\; y^M_m)} 
\nonumber\\ 
\psi^N_{-,m}(x) &=& 
2\frac{\sin(2\pi \; M y^M_m)\;\sin(\pi\; M\;y^M_m)}{\sin(\pi\; y^M_m)} 
=4
\frac{\cos(\pi\; M\;y^M_m)\;\sin^2(\pi\; M\; y^M_m)}{\sin(\pi\; y^M_m)} 
\nonumber\\
&=& 2
\frac{-\cos(3\pi\; M\;y^M_m)+\cos(\pi\; M\; y^M_m)}{\sin(\pi\; y^M_m)} 
\ea
Thus we may define two mother wavelets $\psi_+(x)=\sin(\pi\;x)=\phi(x)$
and $\psi_-(x)=\cos(\pi\;x)=\pm \sqrt{1-\phi(x)}$ algebraically related to 
the mother scaling function $\phi$ and can write (\ref{4.25}) as 
translations and rescalings of rational functions of those
\ba \label{4.26} 
\psi^N_{+,m}(x) 
&=& \frac{([D_{3\;M}-D_{M}]\;T^m_{1/M}\;\psi_+)(x)}{(T_{1/M}^m\;\psi_+)(x)}  
\nonumber\\
\psi^N_{-,m}(x) 
&=& -\frac{([D_{3\;M}-D_{M}]\;T^m_{1/M}\;\psi_-)(x)}{(T_{1/M}^m\;\psi_+)(x)}  
\ea

\subsection{Non-compact case}
\label{s4.2}

Recall the following facts about the topologies of position space and momentum
space via the Fourier transform where we denote by $M$ the spatial resolution
of the lattice $x^M_m$ with either $m\in \mathbb{Z}$ or 
$m\in\mathbb{Z}_M=\{0,1,2,..,M\}$ where for $M$ odd we set  
$\hat{\mathbb{Z}}_M=\{-\frac{M-1}{2},..,\frac{M-1}{2}M\}$ 
(c: compact, nc: non-compact, d: discrete, nd: non-discrete (continuous)):
\be \label{6.1}
\begin{array}{ccc}
{\sf space-topology} & {\sf momentum-topology} & {\sf Fourier-kernel}\\
& \\
{\sf nc,~~ nd:}\;\;\mathbb{R} 
&  {\sf nc,~~ nd:}\;\;\mathbb{R}
& e_k(x)=e^{i\;k\;x}
\\
{\sf nc,~~ d:}\;\;\frac{1}{M}\cdot\mathbb{Z} 
&  {\sf c,~~ nd:}\;\;[-M\pi,\;M\pi) 
& e^M_k(m)=e^{i\;k\; x^M_m}
\\
{\sf c,~~ nd:}\;\;[0,1) 
&  {\sf nc,~~ d:}\;\;\mathbb{Z} 
& e_n(x)=e^{2\pi\; i\; n\;x}
\\
{\sf c,~~ d:}\;\;\frac{1}{M}\cdot \mathbb{Z}_M 
&  {\sf c,~~ d:}\;\;\;\hat{\mathbb{Z}}_M
& e^M_n(m)=e^{2\;\pi\;i\;n\;x^M_m} 
\end{array}
\ee
Accordingly, in the non-compact and comact case respectively, 
the space of Schwartz test functions 
is a suitable subspace of $L=L_2(\mathbb{R},dx)$ and $L=L_2([0,1),dx)$
respectively which have momentum support in $2\pi\mathbb{R}$ and
$2\pi\cdot \mathbb{Z}$ respectively. Upon discretising space into cells 
of width $1/M$ the momentum support $\mathbb{R}$ and $\mathbb{Z}$ respectively 
gets confined to the Brillouin zones 
$[-\pi\;M,\pi M)$ and $\mathbb{Z}_M$ respectively.

The corresponding completeness relations or resolutions of the identity 
read
\ba \label{6.2}
\delta_{\mathbb{R}}(x,x') &=& \int_{\mathbb{R}}\;\frac{dk}{2\pi}\; e_k(x-x')
\nonumber\\
M\delta_{\mathbb{Z}}{m,m'} 
&=& \int_{-\pi \;M}^{\pi M}\; \frac{dk}{2\pi}\; e^M_k(m-m')
\nonumber\\
\delta_{[0,1)}(x,x') &=& 
\sum_{n\in \mathbb{Z}}\; e_n(x-x')
\nonumber\\
M\delta_{\mathbb{Z}_M}{m,m'} &=& \sum_{n\in \mathbb{Z}_M}\; e^M_n(m-m')
\ea
While the first and third relation in (\ref{6.2}) define the $\delta$ 
distribution on $\mathbb{R}$ and $[0,1)$ respectively, the second and fourth
relation in (\ref{6.2}) are the restrictions to the lattice of the regular
functions 
\ba \label{6.3}
\delta_{\mathbb{R},M}(x) &=& \int_{-\pi \;M}^{\pi M}\; 
\frac{dk}{2\pi}\; e_k(x)=\frac{\sin(\pi\;M\;x)}{\pi\; x}
\nonumber\\
\delta_{[0,1),M}(x) &=& \sum_{n\in \mathbb{Z}_M}\; e_n(x) 
=\frac{\sin(\pi\;M\;x)}{\sin(\pi\; x)}
\ea
which we recognise as the Shannon (sinc) and Dirichlet kernel respectively. 
These 
kernels can be considered as regularisations of the afore mentioned 
$\delta$ distributions in the sense that the momentum integral 
$k\in \mathbb{R}$ or momentum sum $n\in \mathbb{Z}$ has been confined to
$|k|<\pi M$ and $|n|<\frac{M-1}{2}$ respectively. Both are real valued, 
smooth, strongly peaked at $x=0$ and have compact momentum support. 
The Shannon kernel like the Dirichlet kernel 
is an $L_2$ function but it is not of rapid decay with respect to position.\\   
\\
In the previous subsection we already have explored the Dirichlet kernel and 
proved it to be both very useful for renormalisation and for defining
a generalised MRA respectively. In this section we will show that analogous
properties hold for the Shannon kernel.\\ 
\\
We begin with the MRA structure of the Shannon kernel which we denote 
by $\delta_M(x)$ for the rest of this section. As it involves the 
Fourier modes $|k|\le \pi M$ in complete analogy to the compact case we 
consider the space $V_M$ as the closure in $L=L_2(\mathbb{R},dx)$ of the 
smooth functions with compact momentum support in $(-\pi\; M,\pi \;M)$.
This obviously gives a nested structure of subspaces $V_M\subset V_{M'}$
in fact for any positive real numbers $0<M\le M'$ with the ususal ordering 
relation, but due to the lattice context we restrict again to the odd 
positive integers $\cal M$ with $M\le M'$ if $M'/M\in \mathbb{N}$. 

The analogy to the compact case suggests to consider the functions 
\be \label{6.4}
\chi^M_m(x):=\frac{1}{M}\;\delta_M(x-x^M_m)\in V_M
\ee
We have 
\ba \label{6.5}
&& M^2\; <\chi^M_m,\;\chi^M_{m'}>_L
=
\int_{-\pi M}^{\pi M}\; \frac{dk}{2\pi}
\int_{-\pi M}^{\pi M}\; \frac{dk'}{2\pi} \; e_k(x^M_m)\; e_{-k'}(x^M_{m'})\;
<e_k,\;e_{k'}>_L
\nonumber\\              
&=&
\int_{-\pi M}^{\pi M}\; \frac{dk}{2\pi}\; e_k(x^M_m)\; e_{-k}(x^M_{m'})\;
<e_k,\;e_{k'}>_L
\nonumber\\              
&=& \delta_M(x^M_m-x^M_{m'})=M\;\delta_{m,m'} 
\ea
which shows that the $\chi^M_m$ form an orthogonal system of functions 
in $V_M$. 

Next let $f$ belong to the dense subset of $V_M$ consisting of smooth 
functions with compact momentum support in $(-\pi M,\pi M)$. We have 
\ba \label{6.7}
&& M\sum_{m\in \mathbb{Z}}\;\chi^M_m(x)\;<\chi^M_m,f>_L
=
M^{-1}\int_{-\pi M}^{\pi M}\; \frac{dk}{2\pi}\; \hat{f}(k)\;
\int_{-\pi M}^{\pi M}\; \frac{dk'}{2\pi}\; e_{k'}(x)
[\sum_{m\in \mathbb{Z}}\;e^{i\frac{k-k'}{M}\;m}]
\nonumber\\
&=&
M^{-1}\int_{-\pi M}^{\pi M}\; \frac{dk}{2\pi}\; \hat{f}(k)\;
\int_{-\pi M}^{\pi M}\; \frac{dk'}{2\pi}\; e_{k'}(x)
[2\pi \sum_{m\in \mathbb{Z}}\;\delta_{\mathbb{R}}((k-k')/M-2\pi m)]
\nonumber\\
&=&
\int_{-\pi M}^{\pi M}\; \frac{dk}{2\pi}\; \hat{f}(k)\;e_k(x)=f(x)
\ea
where the Fourier transform of $f$ is 
\be \label{6.8}
\hat{f}(k)=\int_{\mathbb{R}}\; dx\; e_{-k}(x)\; f(x)=<e_k,\;f>_L
\ee
and where we used that for $|k|,|k'|<\pi M$ the condition $k-k'=2\pi m$ 
has a solution only for $m=0$ which is $k'=k$. It follows that 
the $\sqrt{M}\chi^M_m$ form an ONB of $V_M$. We may write them as 
\be \label{6.9}
\sqrt{M}\chi^M_m=M^{1/2}\;D_M\; T_{1/M}^m\;\phi,\; \phi(x)={\rm sinc}(\pi x)
\ee
demonstrating that the scaling function of this MRA is nothing but the 
sinc function. While this is well known, we have rederived this here without
any effort and from the regularisation of $\delta$ function perspective 
which in turn is motivated by the desire to produce MRA onb bases with 
locality features. 

Next we turn to the underlying wavelet structure. Again we proceed in 
complete analogy to the compact case and consider the sequence 
$M_N=3^N,\; N\in \mathbb{N}_0$ (we could also allow $N\in \mathbb{Z}$,
however for the purpose of renormalisation one is interested in large 
$M$ only). 
We thus have to decompose $V_{3M}$ into $V_M$ and its orthogonal 
complement $W_M$ in $V_{3M}$. To do this note that
\be \label{6.10}
\chi^{3M}_m(x)=
[e^{2\pi\;M\;(x-x^{3M}_m)}
+e^{-2\pi\;M\;(x-x^{3M}_m)}
+1]\;\int_{-\pi M}^{\pi M}\; \frac{dk}{2\pi}\; e_k(x-x^{3M}_m)
\ee
The integral that appears in (\ref{6.10}) defines an element of $V_M$
and thus can be decomposed into the $\chi^M_m$. The functions that 
appear in the square bracket lie in the span of 
$1,\sin(2\pi M(x-x^M_m)),\cos(2\pi M(x-x^m-m)$. We conclude that 
the $\chi^{3M}_m$ can be decomposed into the functions 
$\chi^M_{\sigma,m'}$ 
with $\sigma=0,\pm 1$ where
$\chi^M_{0,m'}=\chi^M_{m'}$ and 
\be \label{6.11}
\chi^M_{+,m}(x)=\cos(2\pi M (x-x^M_m))\;\chi^M_m(x),\; 
\chi^M_{-,m}(x)=\sin(2\pi M (x-x^M_m))\;\chi^M_m(x),\; 
\ee
Using that $2\pi M+k>\pi M,\;-2\pi M+k<-\pi $ for $|k|<\pi M$ it is not 
difficult to see by a calculation completely analogous to that of 
section \ref{s4} that the $\chi^M_{\sigma,m}$ are mutually orthogonal. 
Thus $W_M$ is spanned by the $\chi^M_{\pm,m}$ which can be written as 
\be \label{6.12}
\chi^M_{\pm,m}=\pm (D_{3M}-D_M) T_{1/M}^m \psi_\pm,\;
\psi_+(x)={\rm sinc}(\pi x),\;
\psi_-(x)={\rm cosinc}(\pi x)
\ee
exhibiting the two mother wavelets. The fact that we can deal here with 
just linear aggregates of mother wavelets rather than rational or algebraic 
ones is due to the fact that in the non compact case the denominator function 
$\pi x$ scales under dilatation while in the compact case the denominator 
function $\sin(\pi x)$ does not. These wavelets are of course 
well known in the literature (there only one mother wavelet is required 
because the MRA is based on powers of 2 rather than powers of 3 as 
considered here), however the novel point here is, apart from 
using powers of 3 rather than 2, that we have constructed them here 
effortlessly, directly, by elementary means starting from the cut-off 
resolution of the identity point of view and without going through the 
complicated algorithm involving Cohen's condition. Note that (\ref{6.12})
extends naturally from $M$ being positive powers of 3 to negative powers.

The analysis (localisation and height) of the extrema of the Shannon kernel
is even simpler than for the Dirichlet kernel since it is basically 
the function ${\rm sinc}(y),\; y=\pi M x$. The absolute maximum is at 
$y=0$ of height 1, the other extrema have to obey $y=\tan(y)$ whose 
approximate solution is $y=\pi/2+N\pi$ for large $y$ with $N\in \mathbb{N}$ 
w.l.g. (we just consider $y\ge 0$ since the function is symmetric). They
thus take the approximate value $(-1)^N/(\pi/2+N\pi)$ and in contrast to the 
Dirichlet kernel decay as a consequence of large $N$ and not because 
they are suppressed by an order of $M$ which is of course the difference 
between the compact and non-compact situation.

As a final remark we note that the Shannon kernel is symmetric and thus 
has an infinite number of odd polynomial vanishing moments (which of course 
do not converge absolutely). 

\subsection{Translation invariant kernels and discretisation of derivatives}
\label{s4.3}

We close this section with the following observation.
\begin{Theorem}[Theorem 6.1] \label{th6.1} ~\\
Suppose that $\partial_M:=I_M^\dagger \;\partial\; I_M$ is the natural
discrete derivative w.r.t. a coarse graining kernel $I_M\;L_M\to L$ and 
such that 
$[\partial,I_M\;I_M^\dagger]=0$. Then for any measurable function $f$ on 
$\mathbb{R}$ we have
$I_M^\dagger \; f(i\partial)\; I_M=f(i\partial_M)$.
\end{Theorem}
Proof:\\
We have 
\be \label{6.47}
\partial_M^N=I_M^\dagger\; (\partial\; [I_M \; I_M^\dagger]^{N-1} \;
\partial\; I_M
\ee
While $I_M^\dagger I_M=1_{L_M}$ by isometry, $p_M:=I_M \;I_M^\dagger$ is a 
projection in $L$ (onto the subspace $V_M$ of the MRA). Thus, if 
$[\partial,p_M]=0$ we find $\partial_M^N=I_M^\dagger \partial^N I_M$.
The claim then follows from the spectral theorem (functional calculus)
since $i\partial_M$ is self-adjoint because $i\partial$ is.\\
$\Box$\\
\\
To see that both the Shannon and Dirichlet kernel satisfy the assumtion
of the theorem it suffices to remark that they only depend on the difference
$x-y$, i.e. they are translation invariant. More precisely, since the 
$\chi^M_m$ with $m\in \mathbb{Z}$ and $m\in \mathbb{Z}_M$ respectively 
are an ONB of $V_M$ just as are the $e_k,\; |k|\le \pi M$ and 
$e_{2\pi n},\;|n|\le \frac{M-1}{2}$ respectively  
\be \label{6.48}
(p_M f)(x) = \sum_m \chi^M_m(x)\; <\chi^M_m,f> 
=\left\{ \begin{array}{cc}
\int_X\; dy\; [\int_{-\pi M}^{\pi M}\;\frac{dk}{2\pi}\; e_k(x-y)]\; f(y)
& X=\mathbb{R}\\
\int_X\; dy\; [\sum_{|n|\le \frac{M-1}{2}}\; e_{2\pi n}(x-y)]\; f(y)
& X=[0,1)
\end{array}
\right.
\ee
and integration by parts does not lead to boundary terms due to the 
support properties of $f$ or by periodicity respectively.  

Translation invariance of the Shannon and Dirichlet kernel respectively is,
besides smoothness, another important difference with the Haar kernel
\be \label{6.49}
\sum_m\; \chi^M_m(x)\chi^M_m(y)= 
=\sum_m\; 
\chi_{[\frac{m}{M},\frac{m+1}{M})}(x)
\chi_{[\frac{m}{M},\frac{m+1}{M})}(y)
\ee
which is not translation invariant. Therefore in this case 
the flows of $\omega_M$ or $\omega_M^{-1}$ are not simply related by 
$\omega_M = I_M^\dagger \omega I_M,\; 
\omega_M^{-1} = I_M^\dagger \omega^{-1} I_M$ and thus one must define 
$\omega_M$ as the inverse of the covariance $\omega_M^{-1}$. As $M\to\infty$
this difference disappears but at finite $M$ it is present and makes the 
study of the flow with respect to a non-translation invariant kernel much 
more and unnecessarily involved. 

In \cite{17} translation invariance of the Shannon and Dirichlet kernels will
be exploited to show that that the discrete fermion theories on the lattices
labelled by $M$  
they define is manifestly doubler free. The Nielsen Ninomiya theorem 
\cite{14a} is 
evaded because the kernels are merely peaked (quasi-local) but not local 
(compact support). This is mechanism is similar as the 
non-locality provided by 
perfect (blocked from the continuum) actions in the Euclidian 
path integral approach \cite{16}.

\section{Free scalar field renormalisation with Dirichlet flow}
\label{s5}

In this section we repeat some of the computations done in \cite{11a} in
terms of the {\it Haar renormalisation flow} but now using the Dirichlet
kernel which may be called the {\it Dirichlet renormalisation flow}. We 
content ourselves by blocking from the continuum.\\
\\
The covariance of the Gaussian measure of a Klein Gordon field on the 
cylinder with mass $p>0$ is $C=(2\omega)^{-1},\;
\omega=\sqrt{-\Delta+p^2}$ where $\Delta$
is the Laplacian on $T^1$. For the massless
case $p=0$ let $Q^\perp=1\;<1,\;>_L$ be the projection on the zero mode
with orthogonal complement $Q=1_L-Q^\perp$. In this case fix any 
number $\omega_0>0$ and set 
\be \label{5.1}
C=Q^\perp\; (2\omega_0)^{-1}\; Q^\perp
+Q\; (2\omega)^{-1}\; Q
\ee
The first observation is that ($1=1_M$ the constant function equal to unity)
\be \label{5.2}
Q^\perp\; I_M\; f_M
=\sum_{m\in \mathbb{Z}_M}\; f_M(m)\; 1\; <1,\chi^M_m>_L
=[\frac{1}{M}\sum_{m\in \mathbb{Z}_M}\; f_M(m)]\; 1\; 
=<1_M,f_M>_{L_M} \; 1_M=:Q^\perp_M\; f_M
\ee
that is $Q^\perp I_M=Q^\perp_M$. We also set $Q_M:=1_{L_M}-Q^\perp_M$.
Then 
\be \label{5.3}
I_M^\dagger\; Q^\perp \; \omega_0^{-1}\; Q^\perp I_M=
Q_M^\perp\;\omega_0^{-1} \; Q_M^\perp
\ee
while 
\ba \label{5.4}
&& (I_M^\dagger\; Q \; \omega^{-1}\; Q\; I_M\; f_M)(m)
=M\; <\chi^M_m,\;Q\;\omega^{-1}\;Q\; I_M\; f_M>_L
\nonumber\\
&=& M\;\sum_{\hat{m}\in \mathbb{Z}_M}\;f_M(\hat{m})\; 
<\chi^M_m,\;Q\;\omega^{-1}\;Q\; \chi^M_{\hat{m}}>_L
\nonumber\\
&=& \sum_{\hat{m}\in \mathbb{Z}_M}\;f_M(\hat{m})\; 
\sum_{0<|n|<\frac{M-1}{2}} \omega(n)^{-1}\;\;e^M_n(\hat{m})^\ast\;
<\chi^M_m,e_n>_L
\nonumber\\
&=& \frac{1}{M}\;\sum_{\hat{m}\in \mathbb{Z}_M}\;f_M(\hat{m})\; 
\sum_{0<|n|<\frac{M-1}{2}} \omega(n)^{-1}\;\;e^M_n(m-\hat{m})
\nonumber\\
&=& \sum_{0<|n|<\frac{M-1}{2}} \omega(n)^{-1}\;\;e^M_n(m)\;
\hat{f}_M(n)
\nonumber\\
&=& [Q_M \; \omega_M\; Q_M\; f_M)(m)
\ea
where the Fourier transform of $f_M\in L_M$ is defined by 
\be \label{5.5}
\hat{f}_M(n)=<e^M_n,f_M>_{L_M},\; f_M=\sum_{n\in \hat{\mathbb{Z}}_M}\;
e^M_n\;\hat{f}_M(n)
\ee
This shows that we have simply $C_M(n)^{-1}/2=\omega_M(n)=\omega(n)=2\pi |n|$ 
for $0<|n|\le (M-1)/2$
and $C_M(0)^{-1}/2:=\omega_0$. In the massive case simply 
$\omega_M(n)=\omega(n),\; 0\le |n|\le (M-1)/2$.  
This should be contrasted with the rather 
complicated expression for $\omega_M(n)$ given for the Haar flow displayed
in \cite{11a} which involves also all the $\omega(n),\; |n|>(M-1)/2$.
While these 
are sub-dominant for large $M$, they are not decaying rapidly. This is caused
by the discontinuity of the Haar scaling function.

These non rapidly decaying terms can cause convergence problems which 
are artefacts of using discontinuous approximants to actually smooth 
functions $f\in L$. In \cite{11b} we encounter the operator on $L$ 
\be \label{5.6}
K_M:=Q\;[C^{-1}-I_M\; C_M^{-1}\; I_M^\dagger]\;Q
\ee
in a massless theory. We compute its action on  $e_n,\;n\not=0$ 
for the Dirichlet flow
\ba \label{5.7} 
\frac{1}{2}\;K_M \; e_n 
&=&
\omega(n)\;e_n-\sum_{\hat{n}\in\hat{\mathbb{Z}}_M}\; C_M^{-1}(\hat{n})
(Q\; I_M e^M_{\hat{n}})\;<e^M_{\hat{n}},I_M^\dagger \;e_n>_{L_M}
\nonumber\\
&=&
\omega(n)\;e_n-\sum_{\hat{n}\in\hat{\mathbb{Z}}_M}\; C_M^{-1}(\hat{n})
(Q\; I_M e^M_{\hat{n}})\;<I_M\; e^M_{\hat{n}},\;e_n>_L
\nonumber\\
&=&
\omega(n)\;e_n-\sum_{0\not=\hat{n}\in\hat{\mathbb{Z}}_M}\; 
\omega_M^{-1}(\hat{n})\; \delta_{n,\hat{n}}\; e_n
\nonumber\\
&=&
\omega(n)\;\theta(|n|-\frac{M-1}{2}+1) \; e_n
\ea
where $\theta$ is the Heavyside step function and 
where we used for $|n|\le (M-1)/2$ 
\be \label{5.8}
I_M\;e^M_n
=\sum_{m\in \mathbb{Z}_M}\; \chi^M_m\; e^M_n(m)
=\frac{1}{M}\;\sum_{|\hat{n}|\le \frac{M-1}{2}}\;e_{\hat{n}}\; 
\sum_{m\in \mathbb{Z}_M}\; e^M_{n-\hat{n}}(m)
=e_n
\ee
Accordingly
\be \label{5.9}
K_M f=\sum_{|n|>\frac{M-1}{2}}\; \omega(n)\; e_n\; \hat{f}(n)
\ee
where the Fourier transform 
\be \label{5.10}
\hat{f}(n)=<e_n,f>_L,\;
f=\sum_{n\in \mathbb{Z}}\; \hat{f}_n\; e_n
\ee
was used. For smooth $f$, $\hat{f}(n)$ is of rapid decay thus we get the 
sup norm estimate
\be \label{5.11}
||K_M f||_\infty\;\le\;2\pi\;
\sum_{|n|>\frac{M-1}{2}}\;\frac{|n|}{n^4}\;\;|n^4 \hat{f}(n)|
\le M^{-1}\;[\sum_{|n|>\frac{M-1}{2}}\;n^{-2}]\;
\sup_{n\in \mathbb{Z}}\; |n^4 \hat{f}(n)|
\ee
which decays as $M\to \infty$. Thus, since $T^1$ is compact and therefore 
$||.||_1\le ||.||_2\le ||.||\infty$ we get convergence to zero wth respect 
to all three norms. This is not the case with respect to the Haar kernel.

\section{Conclusion and Outlook}
\label{s6}

In this contribution we intended to achieve three goals:\\ 
1.\\ 
To show that a useful (generalised) 
MRA in the compact case can be obtained by direct 
methods not relying on periodisation of non compact MRA's if one is willing 
to accept that the associated ONB is created by rescalings and translations 
of rational rather than linear aggregates of mother scaling functions.
Corresponding mother wavelets then also are to be understood in this 
generalised sense.\\ 
2.\\ 
To show that MRA is directly related to Hamiltonian 
renormalisation and serves as a very useful organisational principle, 
thereby reducing the freedom that one has in choosing the renormalisation 
flow.\\
3. To show that the Dirichlet and Shannon
kernel fits into the generalised MRA scheme
and that its flow has much improved analytical properties as compared 
to the Haar flow.\\
\\
There are many directions into which this work can be extended. We mention 
three of them:\\
I.\\
Instead of the Shannon and 
Dirichlet kernel one can choose other ones that also 
have promising properties. For instance in the compact case 
the Fejer kernel \cite{15}
is the Cesaro average of the Dirichlet kernel, shares many properties 
of the Dirichlet kernel and in addition is manifestly non-negative which 
is not true for the Dirichlet kernel.\\
II.\\
A more general question is: Which kernels are optimal for which 
renormalisation application? Which
ones display as little as possible ``non-linearities'' when one allows 
algebraic rather than linear agrregates of scaling functions to generate 
an MRA? We have seen that translation invariant kernels lead to major
simplifications in the renormalisation flow. See also \cite{13} for 
explicit use of Daubechies wavelets and the properties of the corresponding
flow.\\ 
III.\\
While tori are particularly convenient, there may be other applications 
in which different compact topologies (e.g. spheres) are preferrable. 
We expect that in this case the theory laid out here generalises by 
substituting for the  
the corresponding harmonic analysis (e.g. spherical harmonics $Y_{l,m}$
on $S^2$
rather than toroidal harmonics $e_{n_1,n_2}$ on $T^2$).\\
IV.\\
In the non-compact case the Shannon kernel has a sharp cut-off 
in momentum space at $|k|=\pi M$ of the Fourier transform of the $\delta$
distribution. This causes it to have compact momentum transport but 
to decay only slowly in position space. If we turn the momentum cut-off 
function 
from a step function into a smooth function of compact support or of 
rapid decrease then the 
corresponding kernel will also be of rapid decrease in position space and 
in that sense keep its locality. However, only if it is really of compact
support rather than merely of rapid decrease does it define an MRA 
in the sense of having a nested structure of subspaces of $L$ as otherwise
the spaces $V_M$ all coincide with $L$ (consider e.g a Gaussian cut-off
of width $M$). But even then are simple translates and dilatations of 
the resulting kernel not automatically orthogonal and thus it is not 
a scaling function of the corresponding MRA. In fact it is well known that
no MRA in the strict sense of \cite{5} exists with scaling functions 
of rapid decrease. In that sense the Shannon 
kernel performs better, being simultaneously smooth of some decay and
a scaling function. This suggests that one picks the momentum cut-off not 
sharply (discontinuously) but also not smoothly 
(usually one uses mollifiers based on the 
smooth function $\exp(-x^{-2})$) so that the position 
space decay (locality) is improved while a scaling function results.
For practical calculations it is important
that this momentum cut-off function be analytically managable.



\begin{thebibliography}{99}

\parskip -5pt

\bibitem{1} R. Haag. Local Quantum Physics. Springer Verlag, Berlin,
1984

\bibitem{2} J. Glimm and A. Jaffe. Quantum Physics.
Springer Verlag, New York, 1987

\bibitem{3} 
J. Schwinger. The Theory of Quantized Fields I. 
Physical Review {\bf 82} (1951) 914–927.\\
J. Schwinger. The Theory of Quantized Fields II.\\ 
Physical Review {\bf 91} (1953): 713–728\\
W. Thirring. A Soluble Relativistic Field Theory? 
Annals of Physics. {\bf 3} (1958) 91–112. \\
B. Simon. $P(\Phi)_2$ Euclidian (Quantum) Field Theory. Princeton Series
in Physics, Princeton, (1974) 2016.\\
D. J. Gross, A. Neveu. Dynamical symmetry breaking in asymptotically 
free field theories. Phys. Rev. {\bf D. 10} (1974) 3235–3253\\
J. Glimm. Boson fields with the $:\Phi^4:$ interaction in three dimensions.
Comm. Math. Phys. {\bf 10} 91968) 1


\bibitem{4} K. G. Wilson. The renormalization group: 
Critical phenomena and the Kondo
problem. Rev. Mod. Phys. {\bf 47} (1975) 773

\bibitem{5} C. Chui. An introduction to wavelets. Academic Press, London,
1992

\bibitem{6} A. Haar. Zur Theorie der orthogonalen Funktionensysteme.
Mathematische Annalen {\bf 69} (1910) 331.

\bibitem{7} C. Cattani. Shannon wavelet analysis. Lecture notes in computer 
science {\bf 4488} (2007) 982 

\bibitem{8} I. Daubechies. Orthonormal bases of compactly 
supported wavelets. Comm. Pure and Applied Math. {\bf 41} (1988) 909

\bibitem{9} I. Daubechies. Ten lectures of wavelets. Springer Verlag, Berlin,
1993

\bibitem{10a} A. Cohen, I. Daubechies, P.Vial. Wavelets on the interval and 
fast wavelet transforms. Appl. and Comp. Harm. Anlysis, Elsevier, 1993.
[hal-01311753]

\bibitem{10} B. D. Johnson. A finite dimensional approach to wavelet systems
on the circle. Glasnik Matematicki {\bf 46} (2011) 415.  

\bibitem{11} T. Thiemann. Canonical quantum gravity, constructive QFT and 
renormalisation.
Front. in Phys. {\bf 8} (2020) 548232, Front. in Phys. {\bf 0} (2020) 457.
e-Print: 2003.13622 [gr-qc]

\bibitem{11a} 
T. Lang, K. Liegener, T. Thiemann, Hamiltonian Renormalisation I.
Derivation from Osterwalder-Schrader Reconstruction.
Class. Quant. Grav. {\bf 35} (2018) 245011.
[arXiv:1711.05685];\\
Hamiltonian Renormalisation II.
Renormalisation Flow of 1+1 dimensional free, scalar fields: Derivation.
Class. Quant. Grav. {\bf 35} (2018) 245012.
[arXiv:1711.06727];\\
Hamiltonian Renormalisation III.
Renormalisation Flow of 1+1 dimensional free, scalar fields: Properties.
Class. Quant. Grav. {\bf 35} (2018) 245013.
[arXiv:1711.05688];\\
Hamiltonian Renormalisation IV. Renormalisation Flow of D+1 dimensional
free scalar fields and Rotation Invariance.
Class. Quant. Grav. {\bf 35} (2018) 245014, [arXiv:1711.05695]\\
K. Liegener, T. Thiemann. 
Hamiltonian Renormalisation V. Free Vector Bosons.
Front. Astron. Space Sci. {\bf 7} (2021) 547550. 
Front. Astron. Space Sci. {\bf 0} (2021) 98. e-Print: 2003.13059 [gr-qc]

\bibitem{11b} T. Thiemann, E.-A. Zwicknagel. 
Hamiltonian Renormalisation VI. Parametrised Field Theory on the cylinder.

\bibitem{11c} 
B. Bahr, K. Liegener.
Towards exploring features of Hamiltonian renormalisation relevant for
quantum gravity. Class. Quant. Grav. {\bf 39} (2022) 7, 075010.
e-Print: 2101.02676 [gr-qc]

\bibitem{12} M. Creutz. Quarks, Gluons and Lattices. Cambridge
University Press, Cambridge, 1983

\bibitem{13} P. Federbush. A new formulation and regularization of 
gauge theories using a non-linear wavelet expansion. Comm. Math. Phys. 
{\bf 81} (1981) 327.\\
A. Stottmeister, V. Morinelli, G. Morsella, Y. Tanimoto.
Operator-Algebraic Renormalization and Wavelets.
Phys. Rev. Lett. {\bf 127} (2021) 23, 230601. e-Print: 2002.01442 [math-ph]\\
A. Stottmeister, V. Morinelli, G. Morsella, Y. Tanimoto.
Scaling Limits of Lattice Quantum Fields by Wavelets.
Commun. Math. Phys. {\bf 387} (2021) 1, 299-360. e-Print: 2010.11121 
[math-ph]\\
T. J. Osborne, A. Stottmeister. 
Conformal field theory from lattice 
fermions. e-Print: 2107.13834 [math-ph]\\
T. J. Osborne, A. Stottmeister. Quantum Simulation of Conformal Field Theory
e-Print: 2109.14214 [quant-ph]

\bibitem{13a}
H. Levi. A geometric construction of the Dirichlet kernel. 
Transactions of the New York Academy of Sciences {\bf 36} (1974)
640.

\bibitem{14} Y. Yamasaki. Measures on Infinite Dimensional Spaces.
World Scientific, Singapore, 1985

\bibitem{14a} H.B. Nielsen, M. Ninomiya. A no-go theorem for regularizing 
chiral fermions. Phys. Lett. {\bf B105} (1981): 219–223.

\bibitem{14b} F. A. Berezin. Introduction to Superanalysis. 
D. Reidel Publishing Company, Dordrecht, Holland 1987. 

\bibitem{15} G. Travaglini. Fejer kernels for Fourier series on 
$T^n$ and on compact Lie groups. Mathematische Zeitschrift {\bf 216} (1994)
265

\bibitem{16} W. Bietenholz, U.-J. Wiese. Perfect Lattice Actions for 
Quarks and Gluons. 
Nucl. Phys. {\bf B 464} (1996) 319–352.
[arxiv: hep-lat/95100026]

\bibitem{17} 
T. Thiemann. 
Hamiltonian Renormalisation VII. Free Fermions and doubler free kernels


\end{thebibliography}
\end{document}